\documentclass[prd,twocolumn,aps,showpacs]{revtex4}

\def\[{\left\lbrack}
\def\]{\right\rbrack}

\def\({\left(}
\def\){\right)}

\newcommand{\be}{\begin{equation}}
\newcommand{\ee}{\end{equation}}
\newcommand{\ea}{\end{eqnarray}}
\newcommand{\ba}{\begin{eqnarray}}

\begin{document}

\title{Noncommutativity from the symplectic point of view}
\author{E. M. C. Abreu$^a$\footnote{\sf E-mail: evertonabreu@ufrrj.br}, 
C. Neves$^b$\footnote{\sf E-mail: cneves@fisica.ufjf.br} and W. Oliveira$^b$\footnote{\sf E-mail: wilson@fisica.ufjf.br}} 
\affiliation{${}^{a}$Departamento de F\'{\i}sica, Universidade Federal Rural do Rio de Janeiro\\
BR 465-07, 23890-000, Serop\'edica, Rio de Janeiro, Brazil\\
${}^{b}$Departamento de F\'{\i}sica,
ICE, Universidade Federal de Juiz de Fora,\\
36036-330, Juiz de Fora, MG, Brazil\\
\bigskip
\today}

\begin{abstract}
\noindent
The great deal in noncommutative (NC) field theories started when it was noted that NC spaces naturally arise in string theory with a constant background magnetic field in the presence of $D$-branes. In this work we explore how NC geometry can be introduced into a commutative field theory besides the usual introduction of the Moyal product.  We propose a non-perturbative systematic new way to introduce NC geometry into commutative systems, based mainly on the symplectic approach.  Further, as example, this formalism describes precisely how to obtain a Lagrangian description for the NC version of some systems reproducing well known theories.
\end{abstract}  
\pacs{11.15.-q; 11.10.Ef; 11.10.Nx} 

\maketitle


\section{Introduction}

The main motivation to investigate  NC field theories  started when it was noted that NC spaces naturally arise in string theory 
with a constant background magnetic field in the presence of D-branes \cite{strings}.   

In \cite{banerjee}, R. Banerjee discuss how noncommutative structures appear in planar quantum mechanics providing a useful way of obtaining them.   It was based on the noncommutative algebra in planar quantum mechanics that was originated from 't Hooft's analysis on dissipation and quantization \cite{thooft}.  Banerjee shows precisely in \cite{banerjee} that the noncommutativity in the coordinates or in the momentum are dual descriptions, corresponding to distinct polarizations chosen for converting a second order to a first order system.  The Hamiltonian obtained in \cite{banerjee} following 't Hooft reveals a noncommutative algebra.  Differently, in our work, we will see in detail that following the symplectic formalism the noncommutativity can be introduced naturally.

It is opportune to mention here that this noncommutativity in the context of the string theory with a constant background magnetic field in the presence of D-branes was eliminated constructing a mechanical system which reproduces classical dynamics of the string \cite{string}.   NC field theories have been studied extensively in many branches of physics \cite{witten,other,alexei,belov,szabo,omer,RB,hp}.

The perturbative study of scalar field theories was performed in reference \cite{mrs}.  There, the authors analyzed the IR and UV divergencies and verified that the Planck's constant enters via loop expansion.  Here, differently, we make a non-perturbative approach and we will see that the Planck's constant enters naturally in the theory via Moyal product.

In \cite{DJEMAI1}, a general $\alpha$-deformation of the algebra of classical observables that introduces a general NC quantum mechanics is constructed.  This $\alpha$-deformation is equivalent to some general transformation on the usual quantum phase space variables. 
In other words, the authors discuss the passage from classical mechanics to quantum mechanics and then, to NC quantum mechanics, which allows to obtain the associated NC classical mechanics. This is possible since the quantum mechanics is naturally interpreted as a NC (matrix) symplectic geometry \cite{DJEMAI2}.

In few words we can say that to obtain the NC version of a field theory one replace the usual product of fields in the action by the Moyal product, defined as
\be
\phi_1 (x)\star \phi_2 (x) =exp\left( {i\over 2} \theta^{\mu \nu}
\partial^{x}_{\mu}\partial^{y}_{\nu} \right) \phi_1 (x) \phi_2 (y)\mid_{x=y}, 
\nonumber
\ee
where $\theta^{\mu \nu}$ is a real and antisymmetric constant matrix. As a consequence, NC theories are highly nonlocal. We also note that the Moyal product of two fields in the action is the same as the usual product, provide we discard boundary terms. Thus, the noncommutativity affects just the vertices.

Gauge theories have played an important role in field theories since they are related with fundamental physical interactions in nature. In a more general sense, those theories have gauge symmetries defined by some relations called, in the Dirac's language, first class constraints \cite{PD}. The quantization of these theories demands a special care because of the presence of  gauge symmetries indicating some superfluous degrees of freedom, which must be eliminated before or after  the implementation of a valid quantization process.

On the other hand, the covariant quantization of second class systems is, in general, a difficult task because the Poisson brackets are replaced by Dirac brackets. At the quantum level, the variables become operators and the Dirac brackets become commutators. Due to this,
the canonical quantization process is plagued by serious problems, such as ordering operator problems \cite{order} and anomalies \cite{RR} in the context of nonlinear constrained systems and chiral gauge theories, respectively. In view of this, it seems that it is more natural and safe to develop the quantization of second class systems without invoking Dirac brackets. Actually, it was the strategy followed by many authors over the last decades. The noninvariant system has been embedded in an extended phase space in order to change the second class nature of constraints to first one.   Recently, two of us \cite{ANO} and some other authors \cite{ROTHE} have used the symplectic formalism in order to embed second class systems and properly systematize the symplectic embedding formalism, as done in  BFFT \cite{BT} and iterative \cite{IJMP} methods for example.

In this work we propose a formalism to generalize the quantization by deformation introduced in \cite{DJEMAI1} in order to explore, with a new insight, how the NC geometry can be introduced into a commutative field theory.  To accomplish this, a systematic way to introduce NC geometry into commutative systems, based on the symplectic approach and the Moyal product is presented.  Further, this method describes precisely how to obtain a Lagrangian description for the NC version of the system.  To confirm our approach, we use two well known systems, the chiral oscillator and some nondegenerated classical mechanics.  We computed precisely the NC contributions through this generalized symplectic method and obtain exactly the actions in the NC space found in the literature.   It is important to notice that it is the first part of a formalism which target is to introduce the NC geometry in constrained and non-constrained systems.  Here we introduce the second analysis.

We have organized this paper as follows. In Section II, we introduce our generalized quantization by deformation assuming a generic classical symplectic structure.  To confirm the method, we apply it to the chiral oscillator and to an arbitrary nondegenerated mechanics in section III.  The conclusions and perspectives, as usual, are depicted in the last section.

\section{The NC Generalized Symplectic Formalism}

The quantization by deformation \cite{MOYAL1} consists in the substitution of the canonical quantization process by the algebra ${\cal A}_\hbar$ of quantum observables generated by the same classical one obeying the Moyal product, {\it i.e.}, the canonical quantization
\be
\label{00000}
\lbrace h, g\rbrace_{PB} = \frac{\partial h}{\partial \zeta_a} \omega_{ab} \frac{\partial g}{\partial \zeta_b}  \longrightarrow \frac {1}{\imath\hbar} [{\cal O}_h, {\cal O}_g]\;\;,
\ee

\noindent with $\zeta=(q_i,p_i)$, is replaced by the $\hbar$-star deformation of ${\cal A}_0$, given by

\be
\label{00010}
\lbrace h, g\rbrace_{\hbar} = h*_\hbar g - g*_\hbar h\;\;,
\ee
where
\be
\label{00020}
(h *_\hbar g)(\zeta)\,=\,\exp\{\frac{\imath}{2}\hbar \omega_{ab}\partial^a_{(\zeta_1)}\partial^b_{(\zeta_2)}\}h(\zeta_1)g(\zeta_2)|_{\zeta_1=\zeta_2=\zeta}
\;\;,
\ee
with $a,b=1,2,\dots,2N$ and with the following classical symplectic structure
\be
\label{00030}
\omega_{ab} = \pmatrix{0 & \delta_{ij}\cr -\delta_{ji} & 0}\,\,\,{\text with}\,\,\, i,j=1,2,\dots,N\;\;,
\ee
that satisfies the relation below
\be
\label{00040}
\omega^{ab}\omega_{bc} = \delta^a_c\;\;.
\ee

The quantization by deformation can be generalized assuming a generic classical symplectic structure $\Sigma^{ab}$. In this way the internal law will be characterized by $\hbar$ and by another deformation parameter (or more). As a consequence, the $\Sigma$-star deformation of the algebra becomes
\be
\label{00050}
(h *_{\hbar\Sigma} g)(\zeta)\,=\, \exp\{\frac{\imath}{2}\hbar \Sigma_{ab}\partial^a_{(\zeta_1)}\partial^b_{(\zeta_2)}\}h(\zeta_1)g(\zeta_2)|_{\zeta_1=\zeta_2=\zeta}\,\,,
\ee
with $a,b=1,2,\dots,2N$.

This new star-product generalizes the algebra among the symplectic variables  in the following way
\be
\label{00060}
\lbrace h, g\rbrace_{\hbar\Sigma} = \imath\hbar\Sigma_{ab}\;\;.
\ee

In \cite{DJEMAI1,DJEMAI2}, the authors proposed a quantization process in order to passage the NC classical mechanics to the NC quantum mechanics, through the generalized Dirac quantization,
\be
\label{00070}
\lbrace h, g\rbrace_{\Sigma} = \frac{\partial h}{\partial \zeta_a} \Sigma_{ab} \frac{\partial g}{\partial \zeta_b}  \longrightarrow \frac {1}{\imath\hbar} [{\cal O}_h, {\cal O}_g]_{\Sigma}\;\;.
\ee

\noindent The relation above can also be obtained through a particular transformation on the usual classical phase space, namely,
\be
\label{00080}
\zeta^\prime_a = T_{ab} \zeta^b\;\;,
\ee
where the transformation matrix is
\be
\label{00090}
T = \pmatrix { \delta_{ij} & - \frac 12 \theta_{ij} \cr  \frac 12 \beta_{ij}  & \delta_{ij}}\;\;,
\ee
where $\theta_{ij}$ and $\beta_{ij}$ are antisymmetric matrices. As a consequence, the original Hamiltonian becomes
\be
\label{00100}
{\cal H}(\zeta_a) \longrightarrow {\cal H}(\zeta^\prime_a)\;\;,
\ee
where the corresponding symplectic structure is
\be
\label{00110}
\Sigma_{ab} = \pmatrix{\theta_{ij} & \delta_{ij}+\sigma_{ij} \cr -\delta_{ij}-\sigma_{ij} & \beta_{ij}}\;\;,
\ee
with $\sigma_{ij} = - \frac18 [\theta_{ik}\beta_{kj} + \beta_{ik}\theta_{kj}]$. Due to this, the commutator relations look like
\ba
\label{00115}
\[q^\prime_i, q^\prime_j\] &=& \imath\hbar\theta_{ij}\;\;,\nonumber\\
\[q^\prime_i, p^\prime_j\] &=& \imath\hbar (\delta_{ij} + \sigma_{ij})\;\;,\\
\[p^\prime_i, p^\prime_j\] &=& \imath\hbar\beta_{ij}\;\;.\nonumber
\ea

It is clear that in this paper we are only analyzing systems where this symplectic algebra (\ref{00115}) involves only constants.  It is worthwhile to mention that there are systems where the symplectic algebra (\ref{00115}) involves phase space dependent quantities, rather than just constants.  For instance, we can mention the noncommutative Landau problem (for example in \cite{horvathy} and references therein).  
A particle in the noncommutative plane, coupled to a constant magnetic field and an electric potential will possess an algebra similar to (\ref{00115}).  A phase space dependent algebra occurs for a nonconstant magnetic field, as discussed in \cite{horvathy}.   This problem is the object for future analysis of our method.

At this point, it is important to notice that a Lagrangian formulation was not given. Now, we propose a new systematic way to obtain a NC Lagrangian description for a commutative system. In order to achieve our objective, the symplectic structure $\Sigma_{ab}$ must be fixed firstly and, subsequently, the inverse of $\Sigma_{ab}$ must be computed. As a consequence, an interesting problem arise: if there are some constant ({\it Casimir invariants}) in the system, the symplectic structure has a zero-mode, given by the gradient of these {\it Casimir invariants}. Hence, it is not possible to compute the inverse of $\Sigma_{ab}$. However, in Ref.\cite{CNWO} this kind of problem was solved. On the other hand, if $\Sigma_{ab}$ is nonsingular, its inverse can be obtained solving the relation below
\be
\label{00120}
\int \Sigma_{ab}(x,y) \Sigma^{bc}(y,z) d y = \delta_a^c\delta(x-z)\;\;,
\ee

\noindent which generates a set of differential equations, since $\Sigma^{ab}$ is an unknown two-form symplectic tensor obtained from the following first-order Lagrangian
\be
\label{00130}
{\cal L} = A_{\zeta^\prime_a} \dot\zeta^{\prime a} - V(\zeta^\prime_a)\;\;,
\ee
as being
\be
\label{00140}
\Sigma^{ab}(x,y) = \frac {\delta A_{\zeta^\prime_b}(x)}{\delta \zeta^\prime_a(y)} - \frac {\delta A_{\zeta^\prime_a}(x)}{\delta \zeta^\prime_b(y)}\;\;.
\ee

\noindent Due to this, the one-form symplectic tensor, $A_{\zeta^\prime_a}(x)$, can be computed and, subsequently, the Lagrangian description, Eq. (\ref{00130}), is obtained also. In order to compute $A_{\zeta^\prime_a}(x)$, the Eq. (\ref{00120}) and Eq. (\ref{00140}) are used, which generates the following set of differential equations
\ba
\label{00150}
\theta_{ij} B_{jk}(x,y) + \(\delta_{ij}+\sigma_{ij}\)A_{jk}(x,y) &=& \delta_{ik}\delta(x-y)\;\;,\nonumber\\
A_{jk}(x,y)\theta_{ji}  + \(\delta_{ij}+\sigma_{ij}\)C_{jk}(x,y) &=& 0\;\;,\nonumber\\
- \(\delta_{ij} + \sigma_{ij}\)B_{jk}(x,y) + \beta_{ij}A_{jk}(x,y) &=& 0\;\;,\nonumber\\
A_{kj}(x,y)\(\delta_{ji} + \sigma_{ji}\) + \beta_{ij} C_{jk}(x,y) &=& \delta_{ik}\delta(x-y)\;\;,\nonumber \\
&\mbox{}&
\ea
where
\ba
\label{00160}
B_{jk}(x,y) &=& \(\frac {\delta A_{q^\prime_j}(x)}{\delta q^\prime_k(y)} - \frac {\delta A_{q^\prime_k}(x)}{\delta q^\prime_j(y)}\)\;\;,\nonumber\\
A_{jk}(x,y) &=& \(\frac {\delta A_{p^\prime_j}(x)}{\delta q^\prime_k(y)} - \frac {\delta A_{q^\prime_k}(x)}{\delta p^\prime_j(y)}\)\;\;,\nonumber\\
C_{jk}(x,y) &=& \(\frac {\delta A_{p^\prime_j}(x)}{\delta p^\prime_k(y)} - \frac {\delta A_{p^\prime_k}(x)}{\delta p^\prime_j(y)}\)\;\;.
\ea

\noindent From the set of differential equations, Eq. (\ref{00150}), and the equations above, Eq. (\ref{00160}), we compute the quantities $A_{\zeta^\prime_a}(x)$.

\noindent As a consequence, the first-order Lagrangian can be written as

\be
\label{00185}
{\cal L} = A_{\zeta^\prime_a} \dot\zeta^\prime_a - V(\zeta^\prime_a)\;\;.
\ee
Notice that, despite (\ref{00130}) and (\ref{00185}) have the same form, in (\ref{00185}) the $A_{{\zeta'}_a}$ are completely computed through the solution of the system (\ref{00150}).   In both we have a NC version of the theory as a consequence of the deformation in (\ref{00090}) and its corresponding symplectic structure in (\ref{00110}).   This will be clarified through the examples in the next section.

\section{Examples}

In this section, in order to clarify our method, we will use the formalism developed above in two well known mechanical systems.  The first one is the chiral oscillator, which has a close relationship to the Floreanini-Jackiw version of the chiral boson \cite{FJ} through the mapping used in \cite{bazeia}.  The other one is the so-called nondegenerated mechanics \cite{alexei}.   We will show precisely that the results obtained with our formalism coincide with the ones depicted in these both systems, which comproves the effectiveness of the method described in the section before.

\subsection{The chiral oscillator}

Let us consider a bidimensional model which phase space is reduced. Due to this, the symplectic coordinates are given by $\zeta^\prime_a = (q^\prime_i)$, with $a=i=1,2$, and the canonical momenta conjugated to $q^\prime_i$ are not present. With this in mind, the noncommutative algebra given in (\ref{00115}) is comprised only by the first element.  Therefore, following our procedure, the matrix $\Sigma_{ab}$ defined in (\ref{00110}) now has only one element.  Then we consider the symplectic structure as being
\be
\label{00190}
\Sigma_{ij} \,=\, \theta_{ij}\,=\,  \theta\,\epsilon_{ij}\;\;,
\ee
where $\theta$ is the measure of the noncommutativity.  This reduces the set of differential equations, given in Eq. (\ref{00150}), to 
\be
\label{00200}
\frac {\delta A_{q^\prime_j}(x)}{\delta q^\prime_k(y)} - \frac {\delta A_{q^\prime_k}(x)}{\delta q^\prime_j(y)} \,=\, \theta_{ij}^{-1}\,=\,-\, \theta\,\epsilon_{ij}\;\;.
\ee
Notice that the prime is not the spacial derivative, it was defined in (\ref{00080}).

Now it is easy to see that the equation (\ref{00200}) has the following solution,
\be
\label{00210}
A_{q^\prime_i} = - \,\frac 12 \,\theta\,\epsilon_{ij} q^\prime_j\;\;.
\ee

\noindent Substituting (\ref{00210}) in (\ref{00130}), the first-order Lagrangian is given by
\be
\label{00220}
{\cal L} =  - \frac 12 \,\theta\,\epsilon_{ij}\,\dot q^\prime_i\, q^\prime_j - V(q^\prime_j)\;\;.
\ee

We can assume that the symplectic potential is
\be
\label{00230}
 V(q^\prime_j) = \frac{k{q'}_j^2}{2}\;\;.
\ee

\noindent Thus, we have the mechanical version of the FJ chiral boson, namely, the chiral oscillator (CO) \cite{banerjee2}
\be
\label{00240}
{\cal L} =  - \frac 12 \,\theta\,\epsilon_{ij}\,\dot q^\prime_i  q^\prime_j - \frac{k{q^\prime_j}^2}{2}\;\;.
\ee
where different signs in $\theta$ will correspond to different chiralities, similarly as obtained in \cite{banerjee2} (also studied in \cite{banerjee3}).    

To make an analogy of this model with a well known model for the chiral boson ($k=1$) let us make the following map using the relations described in \cite{bazeia} given by,
\ba
\label{00245}
\partial_t \phi &\leftrightarrow& \partial_t {q'}_j\;\;,\nonumber\\
\partial_x \phi &\leftrightarrow & \theta\,\epsilon_{ij} {q'}_j\;\;,
\ea

\noindent and with this map implemented in Eq. (\ref{00240}), it can be seen directly that the FJ chiral boson model \cite{FJ} was obtained.

Although the chiral oscillator was discussed in details in various contexts (for instance, in \cite{banerjee2,banerjee3} and the references therein), the purpose of this specific example is only to illustrate our method of introducing the noncommutativity via the symplectic method.

\subsection{The nondegenerated mechanics}

With the understanding of preceeding example, it is now easy to see the procedure in a more complicated case, where the $\Sigma_{ab}$ matrix is bigger than before.

In \cite{alexei} it was developed a NC version of an arbitrary nondegenerated mechanical system whose action can be written as 

\be
\label{00266}
S\,=\,\int dt\,L\,(\,q^A,\dot{q}^A\,)\;\;,
\ee
with the configuration space variables $q^A(t),\;A=1,2,\dots,n$ and no constraints in the Hamiltonian formulation.

We consider now the following symplectic structure
\be
\label{00246}
\Sigma_{\alpha\beta} \,=\, \pmatrix{-\,2\,\theta_{ij} & \delta_{ij} \cr -\,\delta_{ji} & 0}\;\;,
\ee
where, from (\ref{00110}), we can see that $\sigma_{ij}=\beta_{ij}=0$.  Using (\ref{00120}) we can construct the following matricial equation,
\be
\label{00247}
\pmatrix{-\,2\,\theta_{il} & \delta_{il} \cr -\,\delta_{il} & 0}\; 
\pmatrix{\Sigma^{q_l\,q_j} & \Sigma^{q_l\,p_j} \cr \Sigma^{p_l\,q_j} & \Sigma^{p_l\,p_j}}\,=\,
\pmatrix{\delta_i^{\;j} & 0 \cr 0 & \delta_i^{\;j}}
\ee

\noindent and, we get
\ba
\label{00248}
-\,2\,\theta_{il} \Sigma^{q_l\,q_j} \,+\, \delta_{il}\,\Sigma^{p_l\,q_j} &=& \delta_i^{\;j}\;\;,\nonumber\\
\delta_{il}\,\Sigma^{q_l\,q_j}  &=& 0\;\;,\nonumber\\
-\,2\,\theta_{il} \Sigma^{q_l\,p_j} \,+\, \delta_{il}\,\Sigma^{p_l\,p_j} &=& 0\;\;,\nonumber\\
-\delta_{il}\,\Sigma^{q_l\,p_j} &=& \delta_i^{\;j}\;\;.
\ea

Solving (\ref{00248}) we have that
\ba
\Sigma^{q_i\,q_j}&=&0\;\;, \nonumber\\
\Sigma^{p_i\,q_j}&=&\delta_{ij}\;\;,\nonumber\\
\Sigma^{p_i\,p_j}&=&-\,2\,\theta_{ij}\;\;.
\ea

Hence
\ba
\label{00150a}
\frac {\delta A_{q_j}(x)}{\delta q_i(y)} \,-\, \frac {\delta A_{q_i}(x)}{\delta q_j(y)} &=& 0\;\;,\nonumber\\
\frac {\delta A_{q_j}(x)}{\delta p_i(y)} \,-\, \frac {\delta A_{p_i}(x)}{\delta q_j(y)} &=& \delta_{ij}\;\;,\nonumber\\
\frac {\delta A_{p_j}(x)}{\delta p_i(y)} \,-\, \frac {\delta A_{p_i}(x)}{\delta p_j(y)} &=& -\,2\,\theta_{ij}\;\;.
\ea

A convenient solution of this system is 
\ba
A_{q_i} \,&=&\, {1\over2}\,p_i\,+\,{1\over2}\,q_i\;\;,\nonumber\\
A_{p_i} \,&=&\, \theta_{im}\,p_m\,-\,{1\over2}\,q_i\;\;.
\ea

Finally, we can construct our first-order Lagrangian as
\ba
\label{00160a}
L &=& \( {1\over2}\,p_i\,+\,{1\over2}\,q_i \)\,\dot{q}_i\,+\(\,\theta_{im}\,p_m\,-\,{1\over2}\,q_i\,\)\,\dot{p}_i\, \nonumber\\ 
&-&\,V\,(q,\dot{q}) \nonumber\\
&=& p_i\,\dot{q}_i\,+\,\dot{p}_i\,\theta_{ij}\,p_j\,-\,V'\,(q,\dot{q})\;\;,
\ea
where $V'\,(q,\dot{q})=V(q,\dot{q})\,+\,{1\over2}q_i\,\dot{q}_i$ and $L$ is the same Lagrangian obtained in \cite{alexei} (in the eq. (4) in \cite{alexei}, the $H_0(q^A,v_A)$ is our $V(q,\dot{q})$). In few words we can say that the quantization of this system takes us to quantum mechanics with the ordinary product substituted by the Moyal product, similarly to the case of a particle on a noncommutative plane \cite{alexei}.

The Lagrangian (\ref{00160a}) is the NC version of the nondegenerated mechanical system described by the Lagrangian $L=L(q_i,\dot{q}_i)$ (\cite{alexei}).   It is easy to see that (\ref{00160a}) has the same number of physical degrees of freedom as the initial system $S$, equation (\ref{00266}).  It can be demonstrated also that the equations of motion of the NC system are the same as for the initial system $S$, modulo the term which is proportional to the parameter $\theta^{AB}$.  Finally, we can say that the configuration space variables have the NC brackets: $\{q^A,q^B\}\,=\,-2\,\theta^{AB}\;\;$ (\cite{alexei}).

\section{conclusions and perspectives}

To deform a system by substituting the classical product by the Moyal product comprises essentially the usual embedding of a commutative system in a NC configuration space.  The final system is now recognized as a NC theory.  The last one has been investigated extenuously in the literature.  

In order to improve the knowledge in non-perturbative processes on how to obtain effectively a NC theory, the authors in \cite{DJEMAI1}  discuss the passage from classical mechanics to quantum mechanics and then to NC quantum mechanics, which allows one to obtain the associated NC classical mechanics.

We believe that in this paper we gave a step further. We have proposed an alternative new way to obtain NC models, based on the symplectic approach. An interesting feature on this formalism lies on the symplectic structure, which is defined at the beginning of the process. The choice of the symplectic structure, subsequently, defines the NC geometry of the model and the Planck's constant enters the theory via Moyal product. This formalism also describes precisely how to obtain a Lagrangian description for the NC version of the system.   

To illustrate our method, we obtained a chiral oscillator \cite{banerjee2,banerjee3} in the NC phase space that is equivalent to the Floreanini-Jackiw chiral boson through a convenient mapping.

Our next target was the NC version of an arbitrary nondegenerated mechanical system which has no constraints in the Hamiltonian formulation and where, now, the configuration space variables have the NC brackets $\{q^A,q^B\}=-2\,\theta^{AB}$. The result coincides with the literature.

It is important to stress that the procedure introduced in this work deals only with non-constrained systems.   A work in progress is the investigation of how NC geometry can be introduced into constrained systems via symplectic approach. We believe that our method will bring new insights in this issue also.  

\section{Acknowledgments}
 
\noindent EMCA would like to thank the hospitality and kindness of the Dept. of Physics of the Federal University of Juiz de Fora where part of this work was done.
The authors would like  to thank CNPq, FAPEMIG and FAPERJ (Brazilian financial agencies) for financial support.

\end{document}